\documentclass[twocolumn,english,aps,prl,tightenlines]{revtex4-1}
\usepackage[T1]{fontenc}
\usepackage[latin9]{inputenc}
\usepackage{amsbsy}
\usepackage{graphicx}
\usepackage{esint}

\makeatletter

\usepackage[pdftex]{xcolor}\usepackage{amsfonts}
\usepackage{mathrsfs}\usepackage{bbm}\usepackage{bbold}\usepackage{subfigure}

\makeatother
\usepackage{pdfpages}
\makeatletter
\AtBeginDocument{\let\LS@rot\@undefined}
\makeatother

\usepackage{babel}
\begin{document}

\title{Smeared nematic quantum phase transitions due to rare-region effects
in inhomogeneous systems}

\author{Tianbai Cui}

\affiliation{School of Physics and Astronomy, University of Minnesota, Minneapolis,
MN 55455, USA}

\author{Rafael M. Fernandes}

\affiliation{School of Physics and Astronomy, University of Minnesota, Minneapolis,
MN 55455, USA}
\begin{abstract}
The concept of a vestigial nematic order emerging from a ``mother''
spin or charge density-wave state has been applied to describe the
phase diagrams of several systems, including unconventional superconductors.
In a perfectly clean system, the two orders appear simultaneously
via a first-order quantum phase transition, implying the absence of
quantum criticality. Here, we investigate how this behavior is affected
by impurity-free droplets that are naturally present in inhomogeneous
systems. Due to their quantum dynamics, finite-size droplets sustain
long-range nematic order but not long-range density-wave order. Interestingly,
rare droplets with moderately large sizes undergo a second-order nematic
transition even before the first-order quantum transition of the clean
system. This gives rise to an extended regime of inhomogeneous nematic
order, which is followed by a density-wave quantum Griffiths phase.
As a result, a smeared quantum nematic transition, separated from
the density-wave quantum transition, emerges in moderately disordered
systems.
\end{abstract}
\maketitle
\textit{Introduction} \textemdash{} In an electronic nematic phase,
the crystalline point group symmetry is lowered by electronic degrees
of freedom \cite{Kivelson1998,Vojta2009,Fernandes2014}. In analogy
to liquid crystals, it can arise via the partial melting of a translational
symmetry-breaking smectic phase, which in electronic systems corresponds
to a spin or charge density-wave (DW). In several materials, the DW
can have multiple wave-vectors $\mathbf{Q}$ related by the symmetries
of the underlying lattice. A well known example is a DW on a square
lattice with possible ordering vectors $\left(Q,0\right)$ and $\left(0,Q\right)$,
related by tetragonal symmetry. In these cases, upon melting the DW,
the system may form a vestigial Ising-nematic phase in which the translational
symmetry of the lattice is preserved, but its rotational symmetry
is broken. In the above example on the square lattice, the nematic
transition lowers the tetragonal symmetry down to orthorhombic.

Such a mechanism for electronic nematicity has been proposed in both
iron-based superconductors \cite{Xu2008,Fang2008,Fernandes2012},
in which the DW is in the spin channel, and in the high-$T_{c}$ cuprates
\cite{Kivelson1998,Wang2014,Schutt2015,Nie2017}, where the DW can
be in both spin and charge channels. Interestingly, both materials
at optimal doping exhibit behavior characteristic of a quantum critical
point, such as strange metalicity and enhancement of the quasiparticle
effective mass \cite{Hashimoto2012,Ramshaw2015}. This has motivated
deeper investigations of quantum nematic phase transitions in metals,
in order to elucidate whether a putative nematic quantum critical
point is a key ingredient of the phase diagrams of these superconducting
compounds \cite{Schattner2016}.

Several theoretical works have shown that, for a perfectly clean quasi-two-dimensional
metallic system, the vestigial Ising-nematic order and the mother
density-wave order undergo a simultaneous $T=0$ first-order transition,
implying the inexistence of a quantum critical point \cite{Qi2009,Millis2010,Kamiya2011,Fernandes2012}.
However, disorder, ubiquitously present in realistic systems, can
have dramatic effects on quantum phase transitions \cite{Neto1998,Sachdev1999,Millis2001,Maebashi2002,Vojta2003,Dobrosavljevic2005,Vojta2005a,Alloul2009,Andersen2010,Andrade2010,Fernandes2014}.
Thus, in order to assess the relevance of nematicity to the properties
of these compounds, it is paramount to understand the interplay between
disorder and nematic order. Previous works have focused mostly on
non-vestigial Ising-nematic phases \cite{Carlson2006,Lee2016,Mishra2016},
and on random-field or random-mass types of disorder in the DW degrees
of freedom \cite{Nie2014,Hoyer2016}. 

\begin{figure}
\begin{centering}
\includegraphics[width=1\columnwidth]{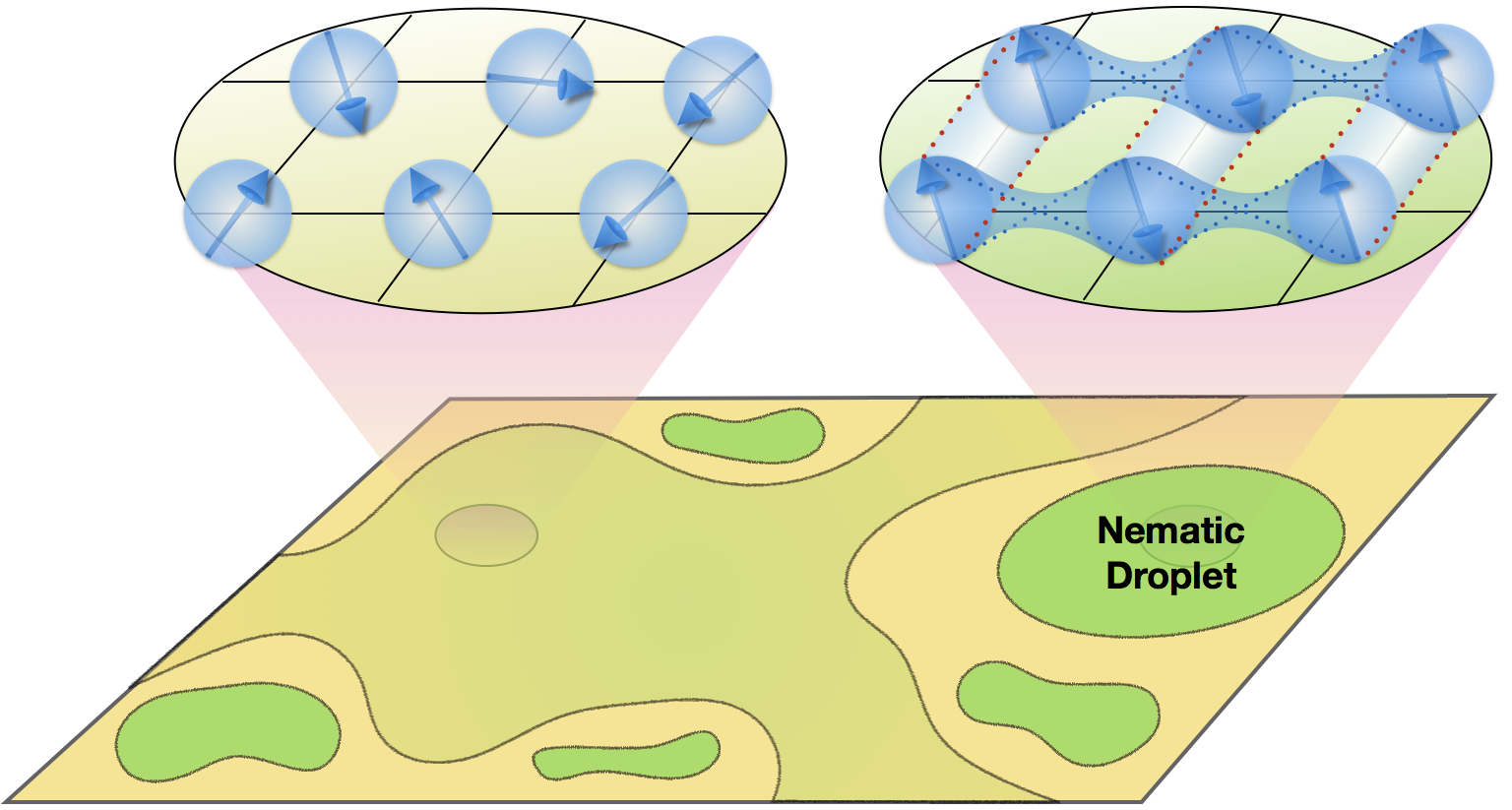}
\par\end{centering}
\caption{Illustration of a moderately diluted system displaying an infinite
cluster (light green) and finite-size droplets (dark green) devoid
of impurities. For this illustration, we consider a ``mother'' spin-density
wave and its vestigial Ising-nematic phase. At $T=0$, finite-size
droplets cannot sustain long-range spin order (indicated by the disordered
spins in the inset), but they can support long-range nematic order
(indicated by the different spin-spin correlations along the $x$
and $y$ axis, resulting in unequal $x$ and $y$ bonds). Importantly,
droplets of moderate sizes undergo a second-order nematic transition
even before the bulk system (and thus the infinite cluster) undergoes
its simultaneous first-order density-wave and nematic transition.
\label{fig_rare_regions}}
\end{figure}

In this paper, we investigate the impact of rare regions on the vestigial
Ising-nematic order arising from a DW quantum phase transition. A
rare region is a relatively large droplet that is devoid of impurities
in a disordered system (see Fig. \ref{fig_rare_regions}). For our
purposes, we consider point-like, randomly diluted impurities that
completely suppress nematic and DW orders locally. Although the probability
of finding such droplets decreases exponentially with their size,
their impact on phase transitions can be significant , causing Griffiths
singularities in thermodynamic quantities \cite{McCoy1968,Griffiths1969}
or smearing phase transitions \cite{Sknepnek2004}. These effects
are particularly strong near a quantum phase transition, due to the
fact that the impurity at $T=0$ is perfectly correlated along the
``time'' axis \cite{Vojta2003,Vojta2005}. 

\begin{figure}
\begin{centering}
\includegraphics[width=1\columnwidth]{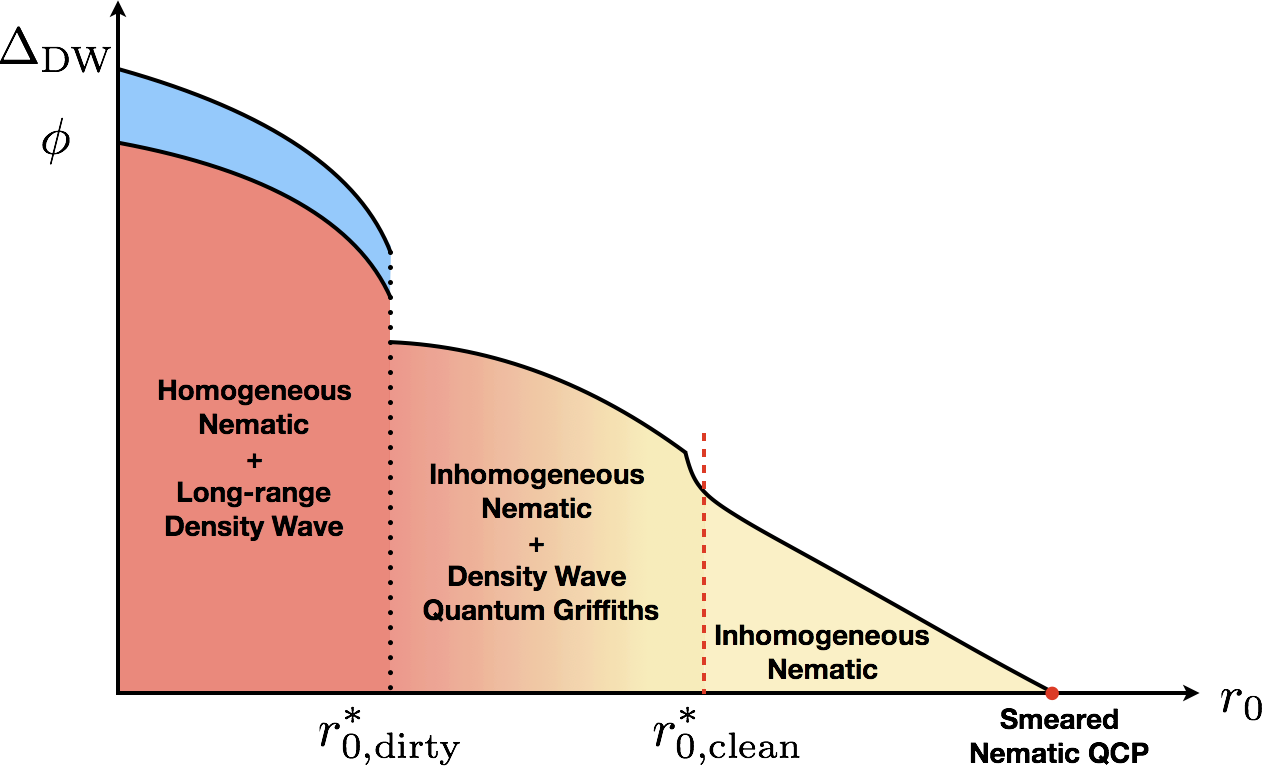}
\par\end{centering}
\caption{Schematic phase diagram illustrating our main results. Here, $r_{0}$
is a control parameter, such as doping or pressure. In the clean system
at $T=0$, the nematic ($\phi$) and density-wave ($\Delta_{\mathrm{DW}}$)
order parameters appear simultaneously at $r_{0,\mathrm{clean}}^{*}$
(dashed line). In the moderately diluted system, which still has a
percolating (infinite) cluster, the first-order quantum transition
is expected to be suppressed down to $r_{0,\mathrm{dirty}}^{*}$ (dotted
line). For $r_{0}>r_{0,\mathrm{clean}}^{*}$, moderately large droplets
undergo a second-order nematic transition, giving rise to inhomogeneous
nematic order. For $r_{0,\mathrm{dirty}}^{*}<r_{0}<r_{0,\mathrm{clean}}^{*}$,
exponentially large droplets have an exponentially large DW correlation
length, resulting in a DW quantum Griffiths phase. Whether $\phi$
jumps or continuously evolves at $r_{0,\mathrm{dirty}}^{*}$ depends
on details of the disorder distribution. \label{fig_PhaseDiagram}}
\end{figure}

As we show here, the rare regions completely change the nature of
the simultaneous first-order nematic-DW quantum transition in a two-dimensional
itinerant system. This is because of the crucial role of the droplet's
dissipative quantum dynamics \cite{Hoyos2006,Hoyos2008,Al-Ali2012},
which allows long-range Ising-nematic order in finite-size droplets
at $T=0$ (see also Ref. \cite{Millis2001}), but not long-range DW
order (for spin or incommensurate charge density-waves). By performing
large-$N$ calculations on a finite-size droplet, we find a wide parameter
range for which the first droplets to order nematically at $T=0$
are not the largest ones, but the droplets of moderately large sizes.
Remarkably, while the \emph{largest droplets} undergo a \emph{first-order}
nematic transition very close to the quantum phase transition of the
clean system, the droplets of \emph{moderate sizes} undergo a \emph{second-order}
nematic transition even before the clean system orders. The result
is the emergence of an inhomogeneously ordered nematic phase, characteristic
of a smeared nematic quantum phase transition \cite{Vojta2003,Hoyos2008},
in the regime where the clean system is not ordered. Our findings,
illustrated in Fig. \ref{fig_PhaseDiagram}, indicate also that a
DW Griffiths phase appears inside this inhomogeneous nematic state,
preceding the onset of long-range DW and homogeneous nematic order.
As we argue below, this behavior may be related to recent puzzling
experimental observations in iron-based compounds.

\textit{Low-energy model} \textemdash{} We consider a general two-dimensional
low-energy model that yields vestigial Ising-nematic order from a
mother DW phase on the square lattice. For concreteness, we consider
two $N$-component DW order parameters, $\boldsymbol{\Delta}_{X}$
and $\boldsymbol{\Delta}_{Y}$, corresponding to two wave-vectors
$\mathbf{Q}_{X}=\left(Q,0\right)$ and $\mathbf{Q}_{Y}=\left(0,Q\right)$
related by tetragonal symmetry. In the case of spin density-wave,
$N=3$ (commensurate) or $N=6$ (incommensurate), whereas for charge
density-wave, $N=1$ (commensurate) or $N=2$ (incommensurate). Hereafter,
we consider only the case $N>1$, as relevant for copper-based and
iron-based superconductors. The low-energy action is given by: \cite{Fernandes2012}
\begin{eqnarray}
S\left[\boldsymbol{\Delta}_{X},\boldsymbol{\Delta}_{Y}\right] & = & \int_{\mathbf{q},\omega}\left[\chi_{\mathbf{q},\omega}^{-1}\left(\boldsymbol{\Delta}_{X}^{2}+\boldsymbol{\Delta}_{Y}^{2}\right)+\frac{u}{2}\left(\boldsymbol{\Delta}_{X}^{2}+\boldsymbol{\Delta}_{Y}^{2}\right)^{2}\right.\nonumber \\
 &  & \left.-\frac{g}{2}\left(\boldsymbol{\Delta}_{X}^{2}-\boldsymbol{\Delta}_{Y}^{2}\right)^{2}\right]\label{action}
\end{eqnarray}
where $\int_{\mathbf{q},\omega}=\int\frac{d^{d}q}{\left(2\pi\right)^{d}}\int\frac{d\omega}{2\pi}$.
Here, $\chi_{\mathbf{q},\omega}^{-1}=r_{0}+\mathbf{q}^{2}+\gamma\left|\omega\right|^{2/z}$
is the inverse DW susceptibility, with the ``tuning parameter''
$r_{0}$ denoting the distance to the mean-field quantum phase transition.
For itinerant systems, the dynamical critical exponent is $z=2$ and
$\gamma$ is the Landau damping coefficient. Vestigial Ising-nematic
order arises when $g>0$. Physically, the emergent Ising-nematic order
parameter $\left\langle \phi\right\rangle =g\left\langle \boldsymbol{\Delta}_{X}^{2}-\boldsymbol{\Delta}_{Y}^{2}\right\rangle $,
which can onset even before the DW, corresponds to unequal fluctuations
around the two DW wave-vectors. Mathematically, it is obtained by
performing a Hubbard-Stratonovich transformation of the second quartic
term of Eq. (\ref{action}):

\begin{equation}
S_{\mathrm{eff}}=\int_{\mathbf{q},\omega}\left\{ \frac{\phi^{2}}{2g}-\frac{\psi^{2}}{2u}+\frac{N}{2}\ln\left[\left(\chi_{\mathbf{q},\omega}^{-1}+\psi\right)^{2}-\phi^{2}\right]\right\} \label{action_eff}
\end{equation}

To obtain this effective action, we also performed a Hubbard-Stratonovich
transformation of the first quartic term of Eq. (\ref{action}) to
introduce the Gaussian-fluctuations field $\left\langle \psi\right\rangle =u\left\langle \boldsymbol{\Delta}_{X}^{2}+\boldsymbol{\Delta}_{Y}^{2}\right\rangle $.
In the large-$N$ limit, and after rescaling the quartic coefficients
$\left(u,g\right)\rightarrow\left(u,g\right)/N$, the equilibrium
values of $\phi$ and $\psi$ as function of $r_{0}$ can be found
within the saddle-point approximation $\frac{\delta S_{\mathrm{eff}}}{\delta\psi}=\frac{\delta S_{\mathrm{eff}}}{\delta\phi}=0$.
Note that $r_{0}+\psi=\left|\phi\right|$ indicates an instability
towards the DW phase, whereas $\phi\neq0$ indicates an instability
towards the Ising-nematic phase. For $d=2$, the Ising-nematic transition
at finite temperatures can be second-order or first-order, but is
split from the DW transition. However, at $T=0$, the two transitions
merge into a single first-order transition (see Fig. \ref{fig_PhaseDiagram}).
These large-$N$ results, reproduced in the Supplementary Material,
were obtained before \cite{Fernandes2012} and confirmed by renormalization
group analysis \cite{Qi2009,Millis2010,Kamiya2011}.

\textit{Large-$N$ solution for a single droplet} \textemdash{} To
assess the relevance of rare regions to the nematic and DW quantum
phase transitions, we first solve the large-$N$ saddle-point equations
for a single droplet of finite linear size $L$, and later average
over the distribution of droplets. The strategy is similar to that
employed in Ref. \cite{Vojta2005} to study Griffiths effects near
a metallic antiferromagnetic quantum critical point. Due to the finite
size of the droplet, the momentum integration in Eq. (\ref{action_eff})
is replaced by a discrete sum over momenta $\mathbf{q}=\frac{2\pi}{L}\mathbf{n}$,
with $\mathbf{n}=\left(n_{x},\,n_{y}\right)$ and $n_{x,y}$ integer.
The saddle-point equations at $T=0$ become (details in the Supplementary
Material):
\begin{eqnarray}
r & = & r_{0}-ur\left(\ln\frac{\Lambda^{2}}{\sqrt{r^{2}-\phi^{2}}}+1-\frac{\phi}{r}\tanh^{-1}\frac{\phi}{r}\right)\nonumber \\
 &  & +\frac{2\pi u}{L^{2}}\left\{ \mathcal{F}\left[\left(r-\phi\right)L^{2}\right]+\mathcal{F}\left[\left(r+\phi\right)L^{2}\right]\right\} \label{eq:CoupledEqR}\\
\phi & = & \phi g\left(\ln\frac{\Lambda^{2}}{\sqrt{r^{2}-\phi^{2}}}+1-\frac{r}{\phi}\tanh^{-1}\frac{\phi}{r}\right)\nonumber \\
 &  & +\frac{2\pi g}{L^{2}}\left\{ \mathcal{F}\left[\left(r-\phi\right)L^{2}\right]-\mathcal{F}\left[\left(r+\phi\right)L^{2}\right]\right\} \label{eq:CoupledEqPhi}
\end{eqnarray}
where $r=r_{0}+\psi$ is proportional to the inverse squared DW correlation
length, $r_{0}\rightarrow r_{0}+u\int_{\mathbf{q},\omega}\frac{1}{q^{2}+\gamma\left|\omega\right|}$
is the renormalized distance to the bare DW quantum critical point,
and $\Lambda$ is the momentum cutoff, hereafter set to be $1/a$
($a$ is the lattice parameter). Note that the quartic coefficients
have been rescaled by $u\rightarrow u/\left(4N\pi^{2}\gamma\right)$
and $g\rightarrow g/\left(4N\pi^{2}\gamma\right)$. The droplet finite-size
effects are encoded in the function $\mathcal{F}(y)=\frac{1}{\pi}\sum\limits _{\mathbf{n\neq0}}\frac{\sqrt{y}}{\left|\mathbf{n}\right|}K_{1}\left(\left|\mathbf{n}\right|\sqrt{y}\right)$,
where $K_{1}\left(x\right)$ is the modified Bessel function of the
second kind. Because $\mathcal{F}\left(y\gg1\right)\sim y^{1/4}\,\mathrm{e}^{-\sqrt{y}}$,
Eqs. (\ref{eq:CoupledEqR})-(\ref{eq:CoupledEqPhi}) recover the saddle-point
expressions for the infinite system $L\rightarrow\infty$. 

To understand how the finite size of the droplet affects the DW and
nematic transitions, we recall that the DW transition takes place
when $r=\left|\phi\right|$. But because $\mathcal{F}\left(y\ll1\right)\sim-\ln y$,
there is no solution to Eqs. (\ref{eq:CoupledEqR})-(\ref{eq:CoupledEqPhi})
with $r=\left|\phi\right|$. This is a consequence of Mermin-Wagner
theorem: at $T=0$, the finite-size droplet has an effective dimensionality
$d_{\mathrm{eff}}=z=2$, which is the lower critical dimension for
the continuous DW order parameter \cite{Vojta2005}. 

\begin{figure}
\begin{centering}
\includegraphics[width=1\columnwidth]{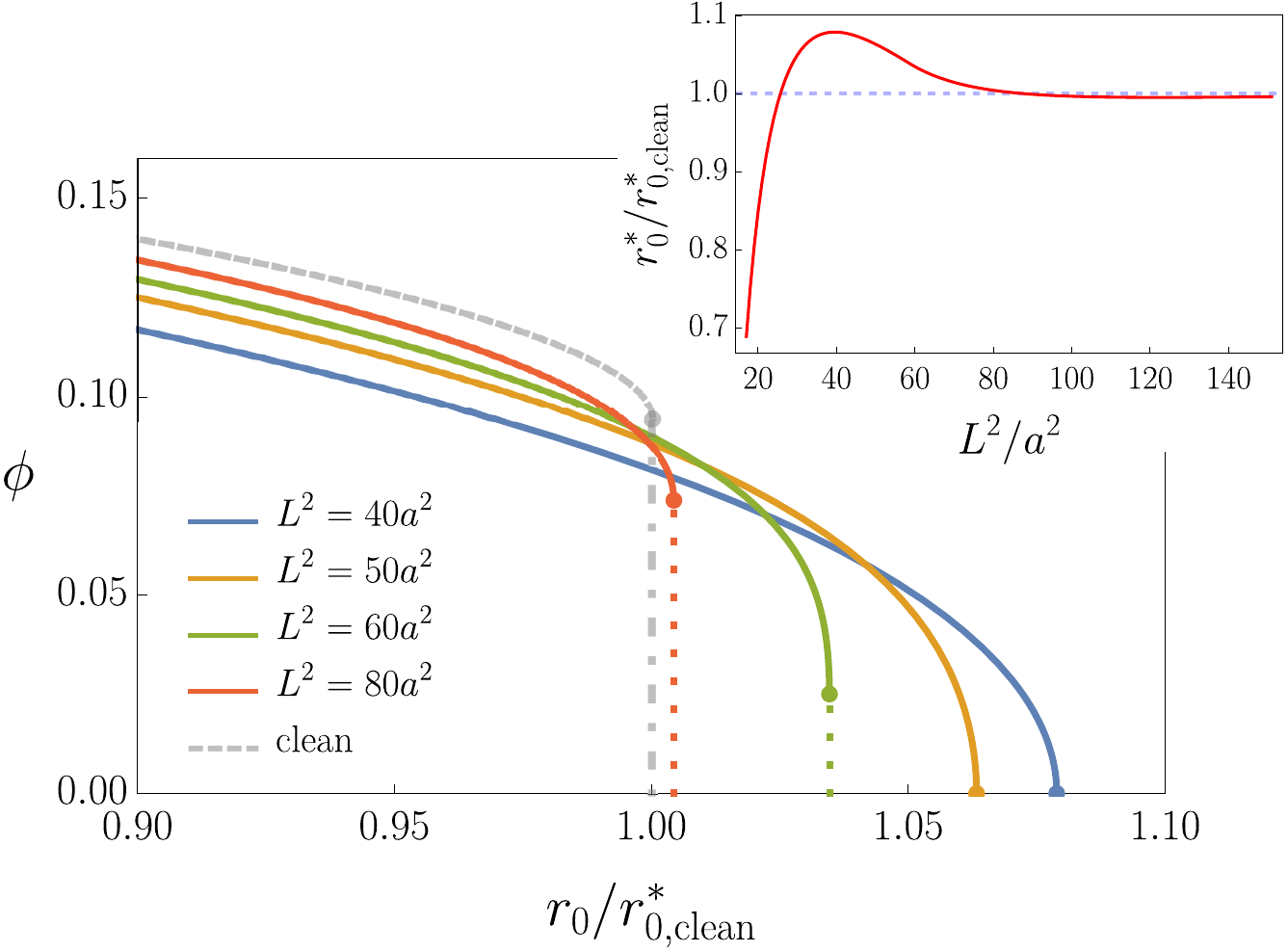}
\par\end{centering}
\caption{Nematic order parameter $\phi$ (in units of $\Lambda^{2}$) inside
a droplet of ``volume'' $L^{2}$, as a function of the control parameter
$r_{0}$ (in units of the value $r_{0,\mathrm{clean}}^{*}$ for which
the clean system undergoes the first-order nematic transition). $\phi_{\mathrm{clean}}\equiv\phi\left(L\rightarrow\infty\right)$
is shown as a dashed line. First-order transitions are indicated by
the dotted line. The inset shows the value of the tuning parameter
$r_{0}^{*}$ at the nematic transition inside a droplet as function
of the droplet volume $L^{2}$. \label{fig_phi}}
\end{figure}

The situation is completely different for the Ising-nematic order
parameter $\phi$: since its lower critical dimension $d_{c}=1<d_{\mathrm{eff}}$,
long-range Ising-nematic order can onset at $T=0$ even in a droplet
of finite size $L$ \cite{Millis2001,Hoyos2008}. To address which
droplets order first, and the character of the Ising-nematic transition
inside them, we numerically solved the coupled Eqs. (\ref{eq:CoupledEqR})-(\ref{eq:CoupledEqPhi})
to obtain $\phi\left(r_{0},L\right)$. The results are shown in Fig.
\ref{fig_phi}; for comparison, we also show the first-order behavior
of the nematic order parameter of the clean system, $\phi_{\mathrm{clean}}\equiv\phi\left(L\rightarrow\infty\right)$,
which orders at $r_{0,\mathrm{clean}}^{*}\equiv r_{0}^{*}\left(L\rightarrow\infty\right)$.

The figure illustrates two very different behaviors: droplets of moderately
large sizes display a non-zero nematic order parameter already in
the non-ordered phase of the clean system, i.e. the nematic transition
inside these droplets happens at $r_{0}^{*}\left(L\right)>r_{0,\mathrm{clean}}^{*}$.
Importantly, the nematic transition in these droplets is generally
second-order. In contrast, large droplets undergo a first-order nematic
transition very close to the clean phase transition, i.e. $r_{0}^{*}\left(L\right)\approx r_{0,\mathrm{clean}}^{*}$.
Note that small droplets (not shown) only order below $r_{0,\mathrm{clean}}^{*}$.
This is more clearly seen in the inset, which shows the nematic transition
parameter $r_{0}^{*}\left(L\right)$ as function of the size $L$.
For the particular values of $u$ and $g$ used here, $u=0.9$ and
$g=0.25u$, the first droplet to order has ``volume'' $L^{2}\approx40a^{2}$,
and all droplets with volumes smaller than $L^{2}=58a^{2}$ undergo
a second-order nematic transition. As we show in the Supplementary
Material, this behavior is not specific to these values of $u$ and
$g$, but happen in a wide region of the $(u,g)$ parameter space.
Importantly, for all droplets, the DW transition does not take place,
i.e. the nematic and DW quantum phase transitions are naturally split
inside a finite-size droplet.

\textit{Average over droplets} \textemdash{} To assess the impact
of the nematically ordered droplets on the thermodynamic properties
of the system, we need to average over the different possible droplets.
The key quantity is thus the probability $P(V)$ of a impurity-free
droplet of volume $V\equiv L^{2}$ being realized in the system, which
in turn is determined by the particular type of disorder distribution.
For concreteness, we consider randomly diluted impurities that kill
DW and nematic order at a given site with probability $1-p$, such
that $p=0$ corresponds to the completely dirty system and $p=1$,
to the completely clean system. Using results of percolation theory
\cite{Stauffer2003}, we can write down the approximate expression
(see Supplementary Material):

\begin{equation}
P(V)=\frac{p_{c}\,V^{1-\tau}\exp\left(-V/V_{0}\right)}{\sum V^{1-\tau}\exp\left(-V/V_{0}\right)}\label{percolation}
\end{equation}

Here, $\tau=187/91$ is a critical exponent, $V_{0}$ is the typical
volume of an impurity-free droplet for a given $p$, and $p_{c}$
is the percolation threshold for clean sites. Because $V_{0}$ changes
from $0$ to $\infty$ from $p=0,\,1$ to the percolation threshold
$p=p_{c}$, we treat $V_{0}$ as our disorder ``tuning parameter,''
instead of $p$. 

We consider here the case where dilution is moderate, and the system
is above the percolating threshold for clean sites, $p>p_{c}$. In
this case, in addition to the finite-size droplets described above,
there is a single infinite percolating droplet devoid of impurities,
which behaves similarly to the clean bulk system (see schematic Fig.
\ref{fig_rare_regions}). Because the infinite droplet has less sites
than the bulk system, the DW-nematic first-order transition inside
of it is expected to happen for $r_{0,\mathrm{dirty}}^{*}<r_{0,\mathrm{clean}}^{*}$.
Thus, for $r_{0}>r_{0,\mathrm{dirty}}^{*}$, the average Ising-nematic
order parameter is given solely by the contributions from the finite-size
droplets, $\bar{\phi}=\sum P\left(V\right)\phi\left(V\right)dV$.
We tacitly assume that there is a very weak inter-droplet interaction
\textendash{} for instance mediated by the lattice \textendash{} that
align the Ising-nematic order parameters of different droplets. 

\begin{figure}
\begin{centering}
\includegraphics[width=1\columnwidth]{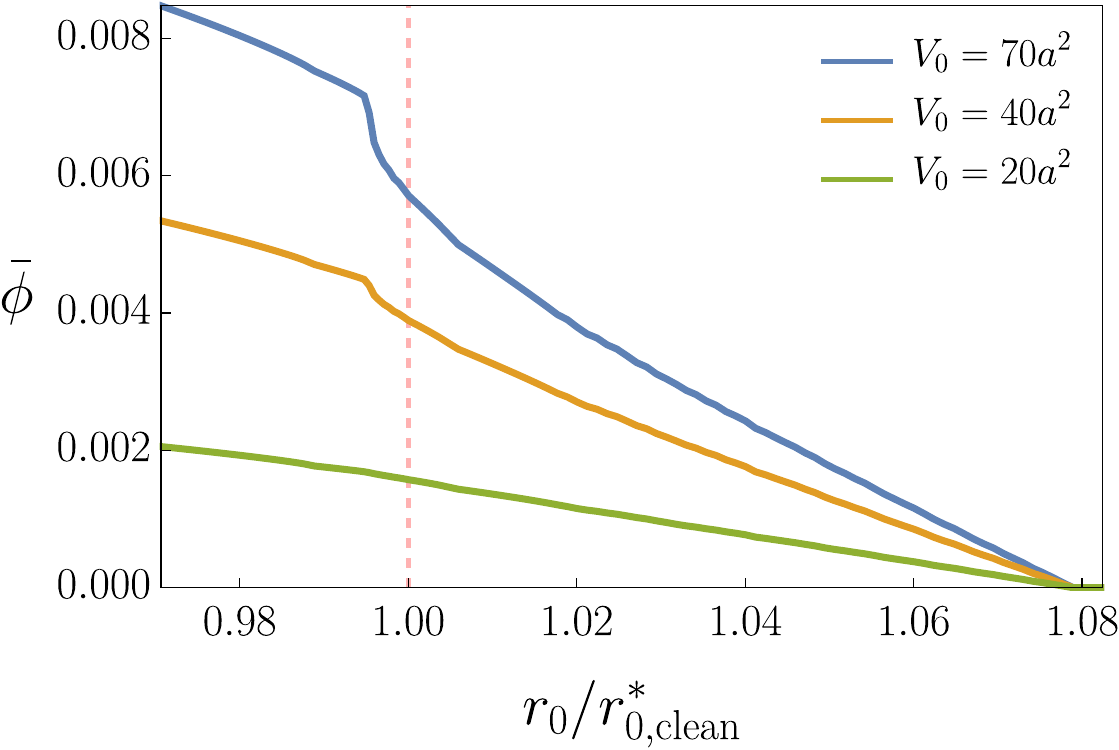}
\par\end{centering}
\caption{Average nematic order parameter of all droplets, $\bar{\phi}$ (in
units of $\Lambda^{2}$), as function of the control parameter, $r_{0}$
(in units of $r_{0,\mathrm{clean}}^{*}$). Different curves correspond
to different ``critical'' droplet volumes $V_{0}$ associated with
the probability distribution (\ref{percolation}). \label{fig_AvePhi}}
\end{figure}

The results for $\bar{\phi}$ are show in Fig. \ref{fig_AvePhi} for
different values of $V_{0}$. Because the first droplets to order
have moderate sizes and undergo a second-order transition, $\bar{\phi}$
seems to evolve continuously as function of $r_{0}$. Our numerical
result suggest a kink of $\bar{\phi}$ at $r_{0,\mathrm{clean}}^{*}$,
although a small jump might also be possible. This behavior is a consequence
of the fact that most of the large droplets order very close to $r_{0,\mathrm{clean}}^{*}$
(see inset of Fig. \ref{fig_phi}). At $r_{0,\mathrm{dirty}}^{*}$,
$\bar{\phi}$ acquires the additional contribution from the infinite
droplet. At this point, the resulting nematic order parameter can
then either undergo a meta-nematic transition, in which it jumps between
two non-zero values, or display another kink in a continuous fashion.
The ultimate behavior is determined by details of the disorder distribution
beyond the scope of our model.

The proliferation of moderate-size droplets sustaining long-range
nematic order upon approaching the clean DW-nematic quantum phase
transition at $r_{0,\mathrm{clean}}^{*}$ signals the emergence of
an inhomogeneously ordered nematic state separated from the DW transition.
The resulting phase transition is thus a smeared quantum phase transition.
In contrast to the more usual smeared quantum phase transitions \cite{Vojta2003,Sknepnek2004,Hoyos2008},
though, it is not driven by exponentially large droplets, and it onsets
already in the regime where the clean system does not order. As for
the DW order, it can only onset at $r_{0,\mathrm{dirty}}^{*}$, since
finite-size droplets cannot sustain DW long-range order. However,
in the regime $r_{0,\mathrm{dirty}}^{*}<r_{0}<r_{0,\mathrm{clean}}^{*}$,
following the arguments of Ref. \cite{Vojta2005}, exponentially large
droplets have an exponentially large correlation length, which promote
quantum Griffiths behavior. Thus, the regime of inhomogeneous nematic
order is followed by a quantum Griffiths DW phase, as shown schematically
in Fig. \ref{fig_PhaseDiagram}. The latter is characterized by power-law
singularities of thermodynamic and DW-related quantities, with non-universal
exponents that depend on $r_{0}$ \cite{Vojta2005}. Note that, although
our analysis has been restricted to $T=0$, we expect the smeared
transition behavior to persist for small enough temperatures, as the
moderate sizes of the relevant droplets can still be smaller than
the nematic correlation length.

\textit{Concluding remarks} \textemdash{} We showed that even weak
disorder fundamentally alters the properties of the Ising-nematic
quantum phase transition associated with a mother charge or spin density-wave
quantum phase transition. The simultaneous first-order transition
of the clean, itinerant system is replaced by an interesting regime
that displays inhomogeneous (but long-range) smeared nematic order
accompanied by a DW quantum Griffiths phase. The extent and relevance
of this regime is controlled by the likelihood of finding isolated
droplets of moderate (rather than very large) sizes, which in turn
is controlled by the strength of disorder.

These results have important implications for the understanding of
the phase diagrams of copper-based and iron-based superconductors,
where an Ising-nematic phase has been argued to emerge from charge
and/or spin density-waves. In the case of the iron pnictides, where
the Ising-nematic and DW transition lines follow each other closely,
these effects are expected to be more pronounced and less ambiguous.
Interestingly, elasto-resistance measurements of the nematic susceptibility
upon approaching the putative nematic-DW quantum phase transition
from high temperatures revealed a weakening of fluctuations and deviation
from Curie-Weiss behavior at low temperatures \cite{Kuo2016}. This
behavior was observed only in compounds with sufficient degree of
disorder. We argue that it could be attributed, at least in part,
to the onset of long-range nematic order in finite-size droplets.
This phenomenon may also help understand the appearance of local inhomogeneous
nematic order in NQR measurements in the nominally tetragonal state
\cite{Lang2010,Dioguardi2016}. Finally, magnetic measurements in
Mn-doped BaFe$_{2}$As$_{2}$, which is significantly less homogeneous
than other doped compounds, have been interpreted in terms of a magnetic
Griffiths phase \cite{Inosov2013,Gastiasoro2014}. It would be interesting
to probe whether local nematic order also emerges in these compounds,
simultaneously to the appearance of the reported Griffiths behavior.
\begin{acknowledgments}
We thank A. Chubukov, P. Orth, and J. Schmalian for fruitful discussions.
This work was supported by the U.S. Department of Energy, Office of
Science, Basic Energy Sciences, under Award number DE-SC0012336.
\end{acknowledgments}

\bibliographystyle{apsrev4-1}
\bibliography{paper.bib}

\pagebreak


\section*{Supplementary material}

# Supplementary Material: "Smeared nematic quantum phase transitions due to rare-region effects in inhomogeneous systems"

Tianbai Cui and Rafael M. Fernandes

*School of Physics and Astronomy, University of Minnesota, Minneapolis, MN 55455, USA*

## I. DERIVATION OF THE SADDLE-POINT EQUATIONS FOR A DROPLET

We start by rewriting the effective large-$N$ rescaled action for a droplet with linear size $L$, as given in the main text:

$$S_{\text{eff}}[\psi, \phi] = \frac{1}{L^2} \sum_{\mathbf{q}} \int \frac{d\omega}{2\pi} \left\{ \frac{\phi^2}{2g} - \frac{\psi^2}{2u} + \frac{1}{2} \ln\left[ \left( \chi_{\mathbf{q},\omega}^{-1} + \psi \right)^2 - \phi^2 \right] \right\} \tag{S1}$$

where $\phi$ is the Ising-nematic order parameter and $\psi$ is the DW fluctuation field. Here, $\chi_{\mathbf{q},\omega} = \left( r_0 + \mathbf{q}^2 + \gamma |\omega| \right)^{-1}$ is the bare DW susceptibility with Landau damping $\gamma$ and $\mathbf{q} = \frac{2\pi}{L}\mathbf{n}$ is the discretized momentum with $\mathbf{n} = (n_x, n_y)$ and $n_x, n_y \in \mathbb{Z}$. The saddle-point of the effective action is given by setting $\frac{\delta S_{\text{eff}}}{\delta \psi} = \frac{\delta S_{\text{eff}}}{\delta \phi} = 0$, which corresponds to the following coupled equations,

$$\frac{r - r_0}{u} = \frac{1}{2L^2} \sum_{\mathbf{q}} \int \frac{d\omega}{2\pi} \left( \frac{1}{r + q^2 + \gamma |\omega| - \phi} + \frac{1}{r + q^2 + \gamma |\omega| + \phi} \right) \tag{S2}$$

$$\frac{\phi}{g} = \frac{1}{2L^2} \sum_{\mathbf{q}} \int \frac{d\omega}{2\pi} \left( \frac{1}{r + q^2 + \gamma |\omega| - \phi} - \frac{1}{r + q^2 + \gamma |\omega| + \phi} \right) \tag{S3}$$

where $r \equiv r_0 + \psi$. To tackle the discrete momentum summation, we apply the Poisson summation formula [1]:

$$\frac{1}{L^d} \sum_{\mathbf{q}} f(\mathbf{q}) = \int \frac{d^d q}{(2\pi)^d} f(\mathbf{q}) + \sum_{\mathbf{n} \neq \mathbf{0}} \frac{d^d q}{(2\pi)^d} f(\mathbf{q}) e^{i\mathbf{q}\cdot\mathbf{n}L} \tag{S4}$$

yielding:

$$\begin{aligned}
\frac{r - r_0}{u} = \frac{1}{2} \int \frac{d^2 q \, d\omega}{(2\pi)^3} & \left( \frac{1}{r + q^2 + \gamma |\omega| - \phi} + \frac{1}{r + q^2 + \gamma |\omega| + \phi} \right) \\
+ \frac{1}{2} \sum_{\mathbf{n} \neq \mathbf{0}} \int \frac{d^2 q \, d\omega}{(2\pi)^3} & \left( \frac{e^{i\mathbf{q}\cdot\mathbf{n}L}}{r + q^2 + \gamma |\omega| - \phi} + \frac{e^{i\mathbf{q}\cdot\mathbf{n}L}}{r + q^2 + \gamma |\omega| + \phi} \right)
\end{aligned} \tag{S5}$$

$$\begin{aligned}
\frac{\phi}{g} = \frac{1}{2} \int \frac{d^2 q \, d\omega}{(2\pi)^3} & \left( \frac{1}{r + q^2 + \gamma |\omega| - \phi} - \frac{1}{r + q^2 + \gamma |\omega| + \phi} \right) \\
+ \frac{1}{2} \sum_{\mathbf{n} \neq \mathbf{0}} \int \frac{d^2 q \, d\omega}{(2\pi)^3} & \left( \frac{e^{i\mathbf{q}\cdot\mathbf{n}L}}{r + q^2 + \gamma |\omega| - \phi} - \frac{e^{i\mathbf{q}\cdot\mathbf{n}L}}{r + q^2 + \gamma |\omega| + \phi} \right)
\end{aligned} \tag{S6}$$

The integrals appearing in the equations above can be evaluated analytically:

$$\begin{aligned}
\int \frac{d^2 q \, d\omega}{(2\pi)^3} \frac{1}{A + q^2 + \gamma |\omega|} &= \int \frac{d^2 q \, d\omega}{(2\pi)^3} \frac{1}{q^2 + \gamma |\omega|} - A \int \frac{d^2 q \, d\omega}{(2\pi)^3} \frac{1}{(A + q^2 + \gamma |\omega|)(q^2 + \gamma |\omega|)} \\
&= \int \frac{d^2 q \, d\omega}{(2\pi)^3} \frac{1}{q^2 + \gamma |\omega|} - \frac{A}{4\pi^2 \gamma} \left[ \ln\left( \frac{\Lambda^2}{A} \right) + 1 \right]
\end{aligned} \tag{S7}$$

and:

$$\int \frac{d^2q d\omega}{(2\pi)^3} \frac{e^{i\mathbf{q}\cdot\mathbf{n}L}}{A+q^2+\gamma|\omega|} = \frac{1}{4\pi^2\gamma} \int d\omega\, K_0\left(\sqrt{A+\omega}L|\mathbf{n}|\right) = \frac{2\sqrt{A}}{L|\mathbf{n}|} K_1\left(\sqrt{A}L|\mathbf{n}|\right) \tag{S8}$$

where we used the fact that the momentum cutoff is such that $\Lambda^2 \gg r \pm \phi$. We verified that this condition is met for the parameters $u$ and $g$ used in the main text. Defining $\tilde{r}_0 = r_0 + u\int \frac{d^2q d\omega}{(2\pi)^3} \frac{1}{q^2+\gamma|\omega|}$, $\tilde{u} = u/\left(4\pi^2\gamma\right)$, and $\tilde{g} = g/\left(4\pi^2\gamma\right)$, we arrive at the equations displayed in the main text:

$$r = \tilde{r}_0 - \tilde{u}r\left(\ln\frac{\Lambda^2}{\sqrt{r^2-\phi^2}} + 1 - \frac{\phi}{r}\tanh^{-1}\frac{\phi}{r}\right) + \frac{2\pi\tilde{u}}{L^2}\left(\mathcal{F}\left[(r-\phi)L^2\right] + \mathcal{F}\left[(r+\phi)L^2\right]\right) \tag{S9}$$

$$\phi = \phi\tilde{g}\left(\ln\frac{\Lambda^2}{\sqrt{r^2-\phi^2}} + 1 - \frac{r}{\phi}\tanh^{-1}\frac{\phi}{r}\right) + \frac{2\pi\tilde{g}}{L^2}\left(\mathcal{F}\left[(r-\phi)L^2\right] - \mathcal{F}\left[(r+\phi)L^2\right]\right) \tag{S10}$$

where we defined the function:

$$\mathcal{F}(y) \equiv \frac{1}{\pi}\sum_{\mathbf{n}\neq\mathbf{0}} \frac{\sqrt{y}}{|\mathbf{n}|} K_1\left(|\mathbf{n}|\sqrt{y}\right) \tag{S11}$$

Note that, for $\phi = 0$, we recover the self-consistent equations of Ref. [1] for an itinerant antiferromagnet. Using the fact that:

$$\frac{\sqrt{y}}{|\mathbf{n}|} K_1\left(|\mathbf{n}|\sqrt{y}\right) = \frac{1}{4}\int_0^\infty du\, u^{-2} e^{-yu-\frac{|\mathbf{n}|^2}{4u}} \tag{S12}$$

we can evaluate the sum and express the function as an integral:

$$\mathcal{F}(y) = \frac{1}{4\pi}\int_0^\infty du\, u^{-2} e^{-yu}\left[\theta_3^2\left(0, e^{-\frac{1}{4u}}\right) - 1\right] \tag{S13}$$

where $\theta_3(a,b)$ is the elliptic theta function. The asymptotic behavior of $\mathcal{F}(y)$ follows from Eqs. (S11) and (S13). For $y \gg 1$, the sum in Eq. (S11) is dominated by the $|\mathbf{n}| = 1$ contribution, and we obtain:

$$\mathcal{F}(y \gg 1) \simeq \frac{2y^{1/4}e^{-\sqrt{y}}}{\sqrt{2\pi}} \tag{S14}$$

For $y \ll 1$, it is more convenient to use the integral representation (S13). Following the same steps as Ref. [1], it follows that the function has a logarithmic divergence:

$$\mathcal{F}(y \ll 1) \simeq -\ln\left(\frac{y\,e^{\gamma+1}}{4\pi}\right) \tag{S15}$$

where $\gamma$ is the Euler number. Equations (S9) and (S10) are solved numerically to yield $\phi(r_0, L)$. In the numerical calculation, we use an interpolation function for $\mathcal{F}(y)$ that interpolates between Eq. (S11) with $80 \times 80$ terms in the sum and the asymptotic behavior in Eq. (S15). The resulting function is shown in Fig. S1. We checked that the error with respect to the exact function $\mathcal{F}$ is no more than 2%.

Before proceeding, we rederive the solutions of the self-consistent equations in the clean limit (i.e. $L \to \infty$, which makes $\mathcal{F}(y) \to 0$). In this case, the solution of Eqs. (S9) and (S10) shows that $r_0$ is maximum for $\phi = r$. However, $\phi = r$ implies a DW transition, since the renormalized DW susceptibility is $(r-\phi)^{-1}$. Thus, one has to go back to the action (S1) and consider the possibility of a finite DW order parameter $\Delta$. In the large-$N$ limit, this corresponds to adding the term $\Delta^2(r-\phi)$ in the action (see for instance [2]). Minimizing with respect to $\Delta$, it is clear that only when $\phi = r$ that we obtain $\Delta \neq 0$. In this case, the self-consistent equations become:

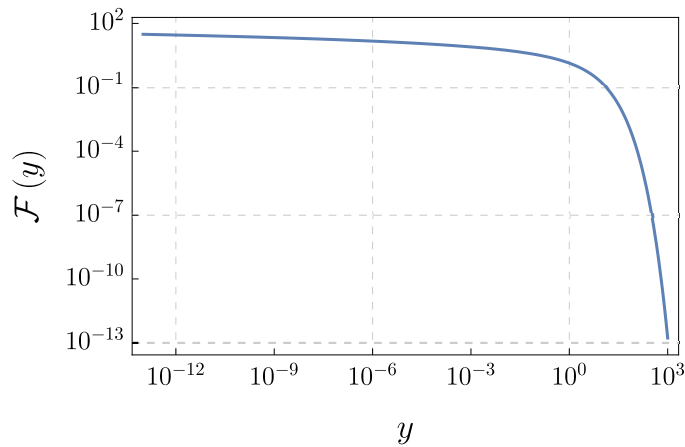

Figure S1. Function $\mathcal{F}(y)$ defined in Eq. (S11).

$$\phi = r_0 - \phi u \left( \ln \frac{\Lambda^2}{2\phi} + 1 \right) + u\Delta^2 \tag{S16}$$

$$\phi = \phi g \left( \ln \frac{\Lambda^2}{2\phi} + 1 \right) + g\Delta^2 \tag{S17}$$

where unimportant constants have been absorbed in $\Delta$. We can now eliminate $\Delta$ to find $r_0$ as function of $\phi$:

$$\frac{r_0}{u} = \phi \left( \frac{1}{u} - \frac{1}{g} \right) + 2\phi \left( \ln \frac{\Lambda^2}{2\phi} + 1 \right) \tag{S18}$$

The first-order transition therefore takes place at:

$$\frac{r_{0,\text{clean}}^*}{\Lambda^2} = u \exp \left( \frac{1}{2u} - \frac{1}{2g} \right) \tag{S19}$$

and the jump in $\phi$ is given by:

$$\phi_{\text{clean}}^* = \frac{r_{0,\text{clean}}^*}{2u} \tag{S20}$$

For the parameters used in the main text, $u = 0.9$ and $g = 0.25u$, we find $\frac{r_{0,\text{clean}}^*}{\Lambda^2} \approx 0.17$ and $\frac{\phi_{\text{clean}}^*}{\Lambda^2} \approx 0.09$. Note that, inside the finite size droplets, $r > \phi$ always. Therefore, we do not need to add the DW field $\Delta$ in the self-consistent equations.

## II. PROPERTIES OF THE ISING-NEMATIC ORDER AS FUNCTION OF THE SIZE OF THE DROPLETS

Here we discuss how two properties of the Ising-nematic order depend on the size $L$ of the droplet: the character of the transition and the value of the tuning parameter $r_0$ for which the quantum transition takes place. We consider here the two self-consistent equations (S9) and (S10) with reduced variables $(r_0, r, \phi) \to (r_0, r, \phi)/\Lambda^2$. Furthermore, the size $L$ of the droplet is measured in units of $1/\Lambda \approx a$, where $a$ is the lattice constant. For convenience, we drop the tilde notation of Eqs. (S9) and (S10) in this section.

The condition for a droplet of size $L$ to undergo a continuous Ising-nematic quantum phase transition is that

$$\left. \frac{dr_0}{d\phi} \right|_{\phi \to 0} \leq 0 \tag{S21}$$

where $r_0$ is the tuning parameter. Taking derivatives with respect to $\phi$ for the self-consistent equations (S9) and (S10), we find:

$$\frac{1}{u}\frac{dr_0}{d\phi} = \left(\frac{1}{u} + \ln\frac{1}{\sqrt{r^2 - \phi^2}}\right)\frac{dr}{d\phi} - u\,\tanh^{-1}\frac{\phi}{r} - 2\pi\left\{\mathcal{F}'\left[(r+\phi)\,L^2\right]\left(1 + \frac{dr}{d\phi}\right) - \mathcal{F}'\left[(r-\phi)\,L^2\right]\left(1 - \frac{dr}{d\phi}\right)\right\} \quad \text{(S22)}$$

$$\frac{dr}{d\phi} = \frac{-\frac{1}{g} + \ln\frac{1}{\sqrt{r^2-\phi^2}} - 2\pi\left(\mathcal{F}'\left[(r+\phi)\,L^2\right] + \mathcal{F}'\left[(r-\phi)\,L^2\right]\right)}{\tanh^{-1}\left(\frac{\phi}{r}\right) + 2\pi\left(\mathcal{F}'\left[(r+\phi)\,L^2\right] - \mathcal{F}'\left[(r-\phi)\,L^2\right]\right)} \quad \text{(S23)}$$

We can eliminate the logarithms by noting that the second self-consistent equation gives:

$$\frac{1}{g} = \ln\frac{1}{\sqrt{r^2 - \phi^2}} + 1 - \frac{r}{\phi}\tanh^{-1}\left(\frac{\phi}{r}\right) - \frac{2\pi}{\phi L^2}\left(\mathcal{F}\left[(r+\phi)\,L^2\right] - \mathcal{F}\left[(r-\phi)\,L^2\right]\right) \quad \text{(S24)}$$

Substituting Eq. (S24) in Eqs. (S22)-(S23), and then expanding for small $\phi$, we find:

$$\left.\frac{dr_0}{d\phi}\right|_{\phi\to 0} = \frac{u}{\lambda}\frac{\phi}{3r^{(0)}}\frac{1 - 4\pi\zeta^2\mathcal{F}'''\left(\zeta\right) - 3\lambda\left[1 + 4\pi\zeta\mathcal{F}''\left(\zeta\right)\right]^2}{1 + 4\pi\zeta\mathcal{F}''\left(\zeta\right)} \quad \text{(S25)}$$

where we defined, for convenience, $\zeta = r^{(0)}L^2$ and $\frac{1}{\lambda} = \frac{1}{u} + \frac{1}{g}$. From (S25), the condition for a droplet to undergo a continuous Ising-nematic quantum phase transition is:

$$(1 - 3\lambda) - 24\pi\lambda\zeta\mathcal{F}''\left(\zeta\right) - 4\pi\zeta^2\mathcal{F}'''\left(\zeta\right) - 48\pi^2\lambda\zeta^2\left[\mathcal{F}''\left(\zeta\right)\right]^2 \leqslant 0 \quad \text{(S26)}$$

Note that, here, $r^{(0)}$ is a shorthand notation for $r^{(0)} \equiv r\left(\phi = 0\right)$. Thus, its value, as well as the corresponding $r_0$, are determined by taking the limit $\phi = 0$ in the self-consistent equations (S9) and (S10):

$$r_0 = ur^{(0)}\left[\frac{1}{u} + 1 - \ln r^{(0)} - \frac{4\pi}{\zeta}\mathcal{F}\left(\zeta\right)\right] \quad \text{(S27)}$$

$$\frac{1}{g} = -\ln r^{(0)} - 4\pi\,\mathcal{F}'\left(\zeta\right) \quad \text{(S28)}$$

Therefore, for given $g$ and $u$, we can find the critical size of the droplet $L_{c1}$ beyond which it undergoes a first-order transition by solving Eqs. (S26) and (S28) simultaneously:

$$L_{c1}^2 = \zeta_{c1}\exp\left[\frac{1}{g} + 4\pi\mathcal{F}'\left(\zeta_{c1}\right)\right] \quad \text{(S29)}$$

where $\zeta_{c1}$ is the value for which Eq. (S26) satisfies the identity. For the parameters used in the main text, $u = 0.9$ and $g/u = 0.25$, we find $L_{c1}^2 \approx 58$ .

Next, we analyze which of the droplets undergoing a second-order phase transition orders at the highest value of the tuning parameter $r_0$, i.e. we look for the critical droplet area $L_{c2}^2$ for which:

$$\frac{dr_0}{dL^2} = 0 \quad \text{(S30)}$$

Taking the derivative with respect to $L^2$ on Eqs. (S27) and (S28), we find:

$$\frac{1}{u}\frac{dr_0}{dL^2} = \frac{dr^{(0)}}{dL^2}\left[\frac{1}{\lambda} + 4\pi\mathcal{F}'\left(r^{(0)}L^2\right)\right] + \frac{4\pi}{L^4}\mathcal{F}\left(r^{(0)}L^2\right) - \frac{4\pi}{L^2}\mathcal{F}'\left(r^{(0)}L^2\right)\left(r^{(0)} + L^2\frac{dr^{(0)}}{dL^2}\right) \quad \text{(S31)}$$

$$0 = -\frac{1}{r^{(0)}}\frac{dr^{(0)}}{dL^2} - 4\pi\mathcal{F}''\left(r^{(0)}L^2\right)\left(r^{(0)} + L^2\frac{dr^{(0)}}{dL^2}\right) \quad \text{(S32)}$$

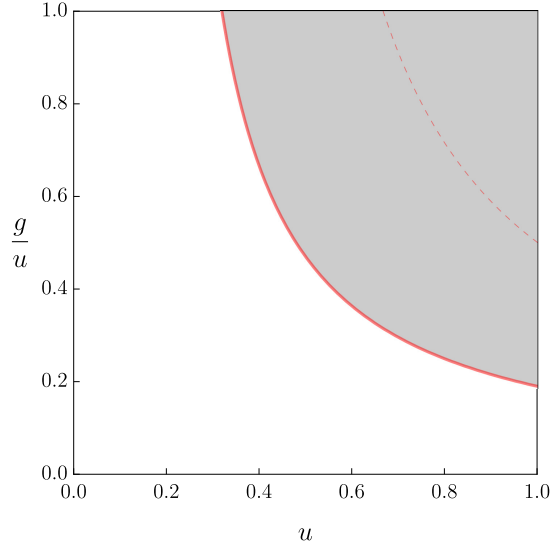

Figure S2. Region in the $(u, g)$ parameter space in which the droplet that orders first does so before the transition of the clean system, i.e, $r_0^* (L_{c2}) > r_{0,\text{clean}}^*$.

Solving for $\frac{dr_0}{dL^2}$, we obtain

$$\frac{dr_0}{dL^2} = \frac{u}{\lambda} \frac{4\pi \exp\left(-\frac{2}{g} - 8\pi \mathcal{F}'(\zeta)\right)}{\zeta + 4\pi \zeta^2 \mathcal{F}''(\zeta)} \left\{ \lambda \left[\mathcal{F}(\zeta) - \zeta \mathcal{F}'(\zeta)\right] \left[\frac{1}{\zeta} + 4\pi \mathcal{F}''(\zeta)\right] - \zeta \mathcal{F}''(\zeta) \right\} \tag{S33}$$

Therefore, the first droplet that undergoes a continuous Ising-nematic quantum phase transition satisfies

$$\left[\mathcal{F}(\zeta_{c2}) - \zeta_{c2} \mathcal{F}'(\zeta_{c2})\right] \left[\frac{1}{\zeta_{c2}} + 4\pi \mathcal{F}''(\zeta_{c2})\right] = \frac{1}{\lambda} \zeta_{c2} \mathcal{F}''(\zeta_{c2}) \tag{S34}$$

and its area is given by:

$$L_{c2}^2 = \zeta_{c2} \exp\left[\frac{1}{g} + 4\pi \mathcal{F}'(\zeta_{c2})\right] \tag{S35}$$

Using Eqs. (S27) and (S28), we can find the corresponding value of the control parameter $r_0^* (L_{c2})$ at which this happens:

$$r_0^* (L_{c2}) = u \exp\left(-\frac{1}{\lambda} - 4\pi \mathcal{F}'(\zeta_{c2})\right) \left[\frac{1}{\lambda} + 1 + 4\pi \mathcal{F}'(\zeta_{c2}) - \frac{4\pi}{\zeta_{c2}} \mathcal{F}(\zeta_{c2})\right] \tag{S36}$$

It is instructive to compare it with the value $r_{0,\text{clean}}^*$ for which the first-order transition of the clean system takes place, see Eq. (S19):

$$\frac{r_0^* (L_{c2})}{r_{0,\text{clean}}^*} = \exp\left(-\frac{1}{2\lambda} - 4\pi \mathcal{F}'(\zeta_{c2})\right) \left[\frac{1}{\lambda} + 1 + 4\pi \mathcal{F}'(\zeta_{c2}) - \frac{4\pi}{\zeta_{c2}} \mathcal{F}(\zeta_{c2})\right] \tag{S37}$$

For the parameters used in the main text, $u = 0.9$ and $g = 0.25u$, we find $L_{c2}^2 \approx 40$ and $\frac{r_0^*(L_{c2})}{r_{0,\text{clean}}^*} \approx 1.08$. More generally, as a function of the two parameters $u$ and $g$, there is a wide region in parameter space in which $\frac{r_0^*(L_{c2})}{r_{0,\text{clean}}^*} > 1$, shown as the shaded area in Fig. S2.

### III.   DERIVATION OF THE PROBABILITY DISTRIBUTION FOR THE DROPLETS SIZES

In this section, we use results from percolation theory to derive the approximate probability distribution function $P(V)$ of finding a finite droplet with "volume" $V = L^2$ that is devoid of impurities. We consider that a local site-impurity completely suppresses both DW and nematic orders. The concentration of impurities is $1 - p$, i.e. the concentration of clean sites is $p$. From percolation theory, the number of finite-size "clean" clusters (droplets) containing $s$ sites, denoted $n_s$, satisfies the scaling relation [3]:

$$n_s = \mathcal{N}_0 s^{-\tau} f(s/s_0) \tag{S38}$$

where $s_0$, the typical cluster size, diverges at the percolation threshold $p_c$, $\tau$ is a critical exponent, and $\mathcal{N}_0$ is a normalization constant. The precise form of the scaling function $f(x)$ can only be obtained numerically. However, an often used approximation is an exponential decay:

$$n_s = \mathcal{N}_0 s^{-\tau} \exp\left(-s/s_0\right) \tag{S39}$$

Now, the total number of clusters of size $s$ is given by $N n_s$, where $N$ is the total number of sites. This implies that the total number of sites that belong to a cluster of size $s$ is $N s n_s$. Therefore, the probability $P(s)$ that a given site belong to a cluster of size $s$ is:

$$P(s) \equiv \frac{N s n_s}{N} = \mathcal{N}_0 s^{1-\tau} \exp\left(-s/s_0\right) \tag{S40}$$

The constant $\mathcal{N}_0$ is given by the normalization condition on $n_s$:

$$\sum_s s n_s = p - P(p) \tag{S41}$$

where $P(p)$ is the number of sites belonging to the infinite cluster (divided by the total number of sites). Thus, we obtain:

$$\mathcal{N}_0 = \frac{p - P(p)}{\sum_s s^{1-\tau} e^{-s/s_0}} \tag{S42}$$

Importantly, $p - P(p)$ vanishes for both $p = 0$ and $p = 1$, and has a maximum near the percolation threshold $p_c$. If we are not too far from the percolation threshold, we can then approximate $p - P(p)$ by $p_c$, as the quantity that changes more strongly upon approaching the percolation threshold is $s_0$, which diverges at $p_c$. We then obtain:

$$P(s) \approx \frac{p_c \, s^{1-\tau} e^{-s/s_0}}{\sum_s s^{1-\tau} e^{-s/s_0}} \tag{S43}$$

The fact that $P(s)$ is not necessarily normalized to 1 is due to the fact that not all sites belong to a finite cluster.

The relationship between $s$ and the linear size $L$ of the cluster that enters the self-consistent equations (S9) and (S10) is $s = L^2/a^2$, where $a$ is the lattice constant. In our problem, as explained above, $L$ is given in units of the inverse momentum cutoff, $L = \tilde{L}/\Lambda$. Thus, we arrive at $\tilde{L}^2 = s(\Lambda a)^2$, with integer $s$. For concreteness, in our calculations we set $\Lambda = 1/a$.

---



\end{document}